\documentclass{aa}
\usepackage[varg]{txfonts}
\usepackage{graphicx}

\begin{document}

\title{A two-component model for fitting light-curves of core-collapse supernovae}
\author{A. P. Nagy\inst{1}
\and J. Vink\'o \inst{1,2}}

\institute{Department of Optics and Quantum Electronics, University of Szeged, Hungary
\and Department of Astronomy, University of Texas at Austin, Austin, TX, USA}

\date{Accepted February 10, 2016}

\titlerunning{Light curve model of core-collapse supernovae}
\authorrunning{A. P. Nagy \& J. Vink\'o}

\abstract{We present an improved version of a light curve  model, which is able to estimate the physical properties of different types of core-collapse supernovae having double-peaked light curves, in a quick and efficient way. The model is based on a two-component configuration consisting of a dense, inner region and an extended, low-mass envelope. Using this configuration, we estimate the initial parameters of the progenitor via fitting the shape of the quasi-bolometric light curves of 10 SNe, including Type IIP and IIb events, with model light curves. In each case we compare the fitting results with available hydrodynamic calculations, and also match the derived expansion velocities with the observed ones. Furthermore, we also compare our calculations with hydrodynamic models derived by the SNEC code, and examine the uncertainties of the estimated physical parameters caused by the assumption of constant opacity and the inaccurate knowledge of the moment of explosion. 
}

\maketitle 

\section{Introduction}
Core-collapse supernovae (CCSNe) form a heterogeneous group of supernova explosion events, but all of them are believed to arise from the death of massive stars ($M$ > 8 $\mathrm{M_\odot}$). The classification of these events is based on both their spectral features and light curve properties. Core-collapse SNe are divided into  se\-veral groups, namely: Type Ib/Ic, Type IIP, Type IIb, Type IIL, and Type IIn \citep{Filippenko}. The different types of core-collapse SNe are thought to represent the explosion of stars with different progenitor properties, such as radii, ejected mass and mass-loss \citep[e.g.,][]{Heger}. Mass-loss may be a key parameter in determining the type of the SN: stars having larger initial masses tend to lose their H-rich envelope, leading to Type IIb, or Type Ib/Ic events, unlike the lower mass progenitors which produce Type IIP SNe. On the other hand, interaction with binary companion may also play an important role in determining He appearance of the explosion event \citep[e.g.,][]{Podsiadlowski, Eldridge, Smartt09}.

Type IIP supernovae (SNe) are known as the most common among core-collapse
SN events. This was recently revealed by \cite{Smith} based on the data from the targeted Lick Observatory Supernova Search \citep[LOSS;][]{Leaman}, and also by \cite{Arcavi10} who used data from the
untargeted Palomar Transient Factory \citep[PTF;][]{Rau} survey. They probably originate from a red supergiant (RSG) progenitor star \citep[e.g.,][]{Grassberg,GraNad76,Chugai, Moriya}. The light curves of Type IIP SNe  are characterized by a plateau phase with a duration of about 80-120 days \citep[e.g.,][]{Hamuy,Dessart,Arcavi} caused by hydrogen recombination, and a quasi-exponential tail determined by the radioactive decay of $^{56}$Co \citep[e.g.,][]{Arnett80,Nadyozhin,Maguire}.
The Type IIb SNe are transitional objects between Type II and Ib explosions, showing strong H and weak He features shortly after the explosion, but the H features weaken and the He lines get stronger at later phases. The weakness of the H features at late phases can be explained by considerable mass-loss from the progenitor star, which causes the stripping of the outermost layer of the hydrogen envelope just before the explosion. One of the most important characteristics of the light curve of these events is the decline rate at late phases, which is go\-ver\-ned by the radioactive decay of $^{56}$Co and the thermalization  efficiency of the gamma-rays produced by the decay processes. 

The collapse of the iron core generates a shock wave that propagates through the envelope of the progenitor star.
Some core-collapse SNe, especially the Type IIb ones, show double-peaked light curves, where the first peak is thought to be dominated by the adiabatic cooling of the shock-heated hydrogen-rich envelope, and the second peak is powered by the radioactive decay of $^{56}$Ni and $^{56}$Co \citep[e.g.,][]{Nakar1}. The double-peak structure may be explained by assuming a progenitor with an extended, low-mass envelope which is ejected just before the explosion \citep{Woosley}. Thus the observed LC of such SNe is generally modeled by a two-component ejecta confi\-guration: a compact, dense core and a more extended, low-mass outer envelope on top of the core \citep{Bersten, Kumar}.  

In this paper we use a semi-analytic light curve model, which is originally presented by \cite{Arnett} and later extended by \cite{Popov93,Blinnikov93} and \cite{Nagy}, to describe the double-peaked light curves of several types of CCSNe. This model is able to produce a wide variety of SN light curves depending on the choice of the initial parameters, such as the ejected mass ($M_{ej}$), the initial radius of the progenitor ($R_0$), the total explosion energy ($E_{tot}$), and the mass of the synthesized $^{56}$Ni ($M_{Ni}$) which directly determines the emitted flux at later phases.    

This paper is organized as follows: in Section 2 we briefly describe the applied light curve model and show the difference caused by various density profile appro\-xi\-mations in the ejecta. In Section 3 we present an estimate of the average opacity from the SNEC hydrodynamic code \citep{Morozova}, and also examine the correlation between the model opacity and ejected mass. Section 4 presents the effect of the uncertainty of the moment of explosion on the derived model parameters. Section 5 and 6 show the application of the two-component configuration for modeling several observed Type IIb and IIP SNe. In Section 7 we compare the expansion velocities derived from the model fits to the observed photospheric velocities of CCSNe. Finally, Section 8 summarizes the main conclusions of this paper.

\section{Two-component light curve model} 

In this paper we generalize the semi-analytic LC model presented by \cite{Nagy}, which assumes a homologously expanding and spherically symmetric SN ejecta. The density structure of the ejecta is assumed to include an inner part with flat (constant) density extending to a dimensionless radius $x_0$, and an outer part where the density decreases as an exponential or a power-law function. Thus, in a comoving coordinate frame the time-dependent density at a particular dimensionless radius ($x$) can be given as
\begin{equation}
\rho(x,t) = \rho(0,0)\ \eta(x) \left(\frac{R_0}{R(t)}\right)^3\ ,
\end{equation}
where $R_0$ is the initial radius of the progenitor, $R(t) = R_0 + v_{exp}\cdot t$ is the radius of the expanding envelope at the given time $t$ since explosion, $v_{exp}$ is the expansion velocity, $\rho(0,0)$ is the initial density of the ejecta at the center ($x$ = 0), and $\eta(x)$ is the dimensionless density profile \citep[see also][]{Arnett}.

The spatial structure of the density for an exponential density profile is
\begin{align}
  \eta(x) = \begin{cases}
      e^{-a (x - x_0)} & \text{if $x>x_0$} \\
      1 & \text{if $x \leq x_0$}\ ,
    \end{cases}
\end{align}
while for the power-law density profile it is
\begin{align}
  \eta(x) = \begin{cases}
      (x/x_0)^{-n} & \text{if $x>x_0$} \\
      1 & \text{if $x \leq x_0$}\ ,
    \end{cases}
\end{align}
where $a$ and $n$ are small positive scalars.  
The initial central density of the ejecta depends on the ejected mass $M_{ej}$ and the 
progenitor radius $R_0$ as
\begin{equation}
\rho(0,0) = \frac{M_{ej}}{4\ \pi\ R_0^3\ f(x_0)}\  ,
\end{equation} 
where $f(x_0)$ is a geometric factor related to the density profile $\eta(x)$ within the ejecta.
For the exponential density profile
\begin{equation}
f(x_0) =\frac{x_0^3}{3} + \frac{1}{a} \left\lbrace x_0^2 +\frac{2}{a} \left[x_0 - e^{(x_0 - 1)}\right] + \frac{2}{a^2} \left[1-e^{(x_0 - 1)}\right] - e^{(x_0 - 1)} \right\rbrace, 
\end{equation}
while for power-law density distribution \citep{Vinko}
\begin{equation}
f(x_0) = \frac{3 x_0^n - n x_0^3}{3 (3-n)} \ .
\end{equation}

The kinetic energy of a model with a given density profile and expansion velocity can be derived as
\begin{equation}
E_{kin} = \frac{1}{2} \int_0^1 4\ \pi\ R_0^3\ \rho(0,0)\ \eta(x)\ x^2\ v(x)^2\ dx\ ,
\end{equation}
where $v(x) = x\cdot v_{exp}$ comes from the condition of homologous expansion. Taking into account
the form of the density profile this integral can be expressed as
\begin{equation}
\begin{split}
E_{kin} & = 2\ \pi\ R_0^3\ \rho(0,0)\ v_{exp}^2\left(\int_0^{x_0} x^4\ dx\ + \int_{x_0}^1 \eta(x)\ x^4\ dx\right)\\ & =  2\ \pi\ R_0^3\ \rho(0,0)\ v_{exp}^2\ g(x_0)\ ,
\end{split}
\end{equation}
where we define $g(x_0)$ as the sum of the two integrals in Eq.~8. 
For the exponential density profile this is
\begin{multline*}
g(x_0)  = \frac{x_0^5}{5}  + \frac{1}{a}\ \lbrace x_0^4 +\frac{4}{a}\ \left[x_0^3 - e^{(x_0 - 1)}\right] + \frac{12}{a^2}\ \left[x_0^2 - e^{(x_0 - 1)}\right] +\\  + \frac{24}{a^3}\ \left[x_0 - e^{(x_0 - 1)}\right] + \frac{24}{a^4}\ \left[1 - e^{(x_0 - 1)}\right] - e^{(x_0 - 1)} \rbrace\ . 
\end{multline*}  
However, for power-law density structure we have
\begin{equation}
g(x_0) = \frac{5 x_0^n - n x_0^5}{5 (5-n)} \ .
\end{equation}
Substituting Eq.~4 into Eq.~7, the expansion velocity turns out to be
\begin{equation}
v_{exp} = \sqrt{\frac{2\ E_{kin}\ f(x_0)}{M_{ej}\ g(x_0)}}\ .
\end{equation} 

Note that we adopt the definition of $v_{exp}$ as the velocity of the outmost layer of the SN
ejecta. This may or may not be related directly to any observable SN velocity. See Sect.~7 for
discussion of the expansion velocities in CCSNe.

In this LC model the energy loss driven by radiation transport is treated by the diffusion approximation, and the bolometric luminosity ($L_{bol}$) is determined by the energy release due to recombination  processes ($L_{rec}$) and radioactive heating ($L_{Ni}$) \citep{Arnett, Nagy}. An alternative energy source, the spin-down of a magnetar \citep{Kasen, Inserra}, is also built-in in the model, which may be useful for fitting the LC of super-luminous SNe (SLSNe). The effect of gamma-ray leakage is also taken into account as $L_{bol} = L_{Ni} (1 - exp(-A_g/t^2)) + L_{rec}$, where the $A_g$ factor refers to the effectiveness of gamma-ray trapping \citep{Chatzopoulos}. It is related to the $T_0$ parameter defined by \cite{Wheeler} as $A_g = T_0^2$.

In the two-component model we use $two$ such spherically symmetric ejecta components, both having different mass, radius, energy and density configuration. The two components have a common center, and
one has much larger initial radius, but smaller mass and lower density than the other. In the 
followings we refer to the bigger, less massive component as the ``envelope'' (or ``shell''), and the more massive, smaller, denser component as the ``core''. Usually, the core has higher kinetic and thermal energy than the outer envelope. This configuration is intended to mimic the structure of a red/yellow supergiant having an extended, low-density outer envelope on top of a more compact and more massive inner region. This configuration is similar to the ejecta model used by \cite{Bersten} for modeling
the LC of the Type IIb SN 2011dh. 
The advantage of this two-component configuration is that it allows the separate solution of the diffusion equation in both components \citep[see][]{Kumar}, if the photon diffusion time-scale 
is much lower in the outer shell than in the core.

For Type IIb SNe we assume that the outer envelope is H-rich, while the inner core is He-rich, both having a constant Thompson-scattering opacity. At early phases the radiation from the cooling envelope dominates the bolometric light curve, while a few days later the LC is governed by the photon diffusion from the inner core, which is centrally heated by the decay of the radioactive nickel and cobalt. The shape of the observed light curve is determined by the sum of these two processes.

Although the plateau-like LCs of Type IIP SNe can be modeled by simple semi-analytic codes 
\citep[e.g.,][]{Arnett89, Arnett, Nagy}, at very early epochs the bolometric LC of 
Type IIP supernovae also shows a faster declining part, which is similar to the first peak in Type IIb LCs. Thus, the two-component ejecta configuration can be a possible solution for modeling the entire LC of Type IIP SNe, but in this case the initial radii and the total energies of the two ejecta
components may have the same order of magnitude. This means that the two components are not 
well-separated, unlike in the Type IIb models. Within this context, the outer envelope may represent the outermost part of the atmosphere of the progenitor star, which has a different density profile and lower mass than the inner region. Note that, an alternative scenario is available in the literature \citep{Moriya, Chugai}. This scheme assumes a low-mass circumstellar medium (CSM) around the progenitor, which may be originated by the mass-loss processes of the star during the RGB phase. The low-mass extended envelope might be  physically associated with this CSM envelope. Indeed, some Type IIP SNe are also reported to show possible effects of CSM interaction in the light curves \citep{Moriya} and also in the spectra \citep{Chugai}. 

\subsection{Temperature profile in the two-component model}
While taking into account the recombination process, the temperature profile may play important role during the calculation of the model LC.    
According to \citet{Arnett}, the temperature distribution within the ejecta is approximated as 
\begin{equation}
T^4(x,t) ~=~ T^4(0,0)\ \psi(x)\ \phi(t) \left(\frac{R_0}{R(t)}\right)^4
\label{temp1}
\end{equation}
where the spatial part, $\psi (x)$, is assumed to be time-independent, and
the function $\phi(t)$ represents a time-dependent scaling factor, while $(R_0/R(t))^4$ describes the adiabatic expansion of the ejecta.
As in \citet{Nagy}, for the spatial profile we assumed Arnett's ``radiative zero''
solution, i.e. 
$\psi(x) ~=~ \sin ( \pi x) / (\pi x)$ for $0 < x < x_i$, where $x_i$ 
is the co-moving dimensionless radius of the recombination front \citep{Arnett, Popov93}. 
This solution
is valid in a constant density ejecta. Since in our models we used a constant-density
configuration for the core component (see below), this temperature profile is a good
approximation for the inner regions. 

\begin{figure*}[!ht]
\resizebox{8cm}{!}{\includegraphics{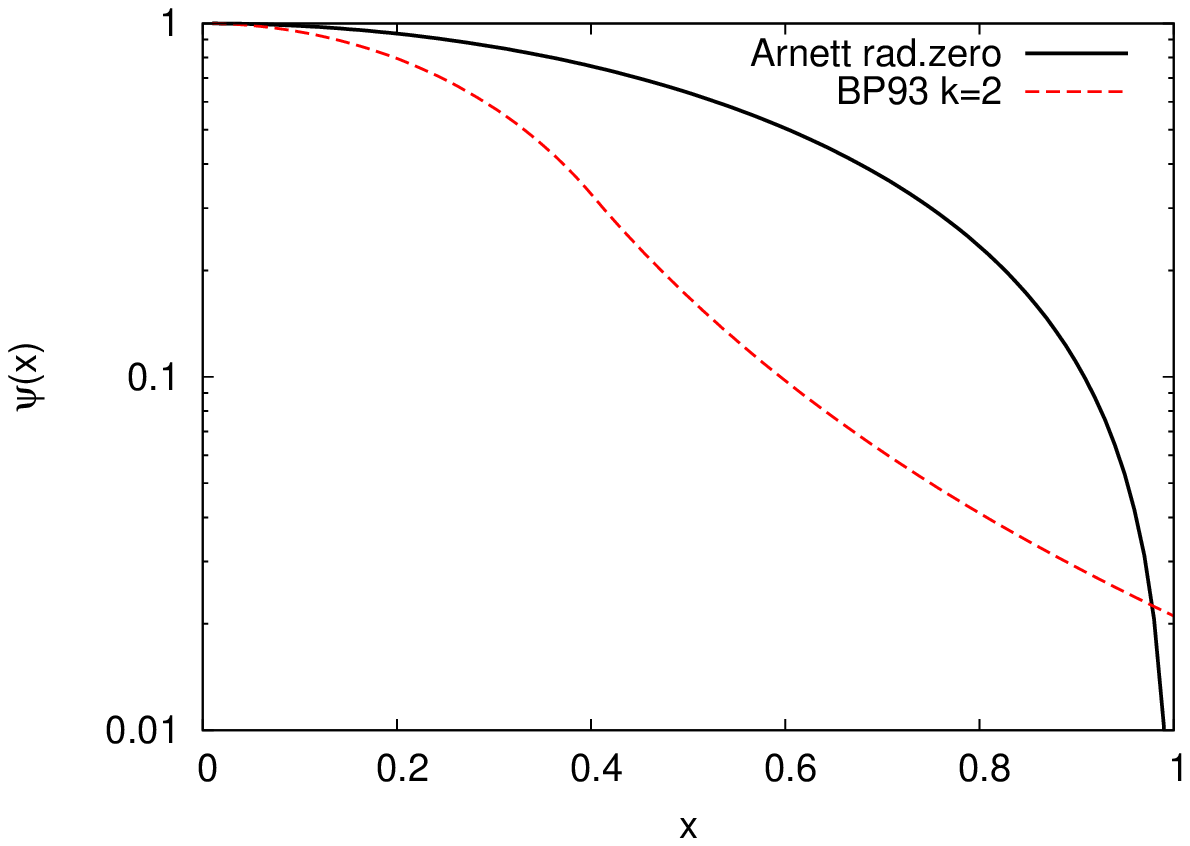}}
\resizebox{8.5cm}{!}{\includegraphics{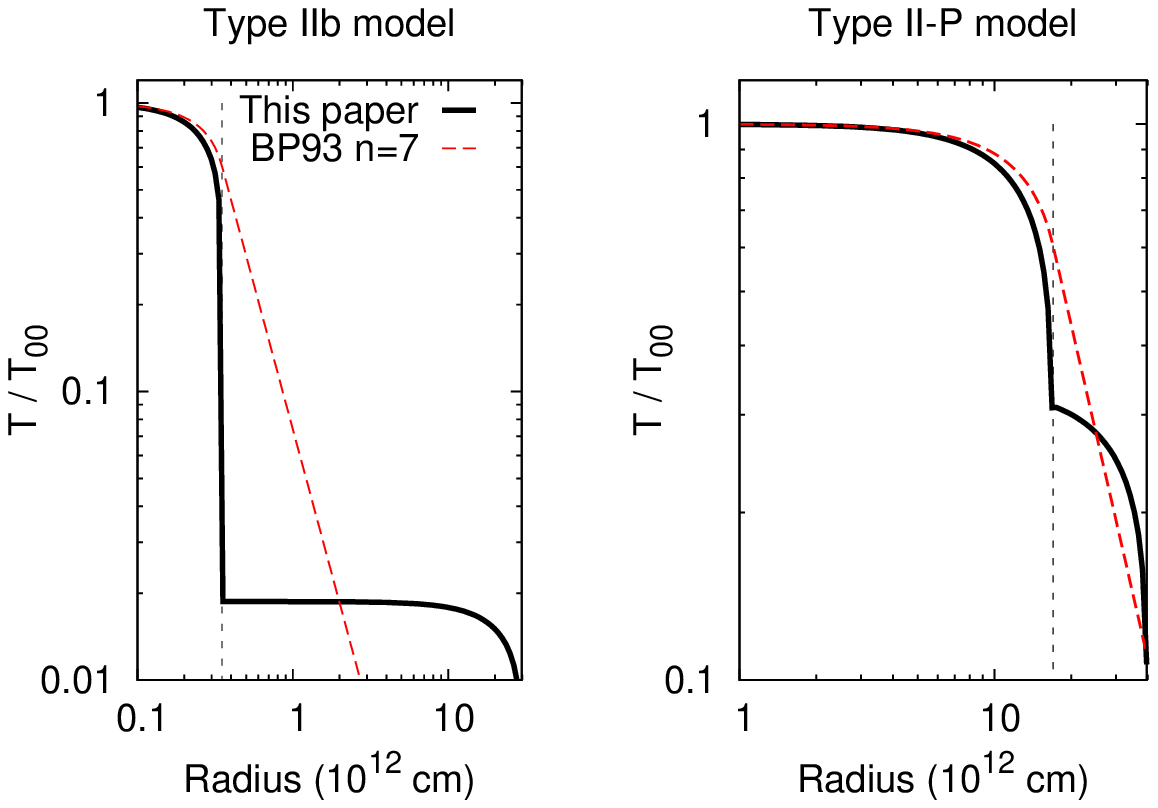}}
\caption{Comparison of temperature profiles applied in our code (continuous lines)
to those of \citet{Blinnikov93} that are valid for power-law ejecta density distribution 
(dashed lines). The left panel shows $\psi(x)$ of Arnett's ``radiative zero'' 
solution together with the BP93 profile for $n=2$ power-law index. In the right panel the 
temperature profiles of our two-component models for both Type IIb and Type II-P
configurations are compared to BP93 profiles having steep ($n=7$) density profiles
on top of the core. See text for explanation.}
\label{fig:temp}
\end{figure*}

This is, however, not necessarily true for the extended envelope, where we applied a 
monotonically decreasing density distribution having a power-law index of $n=2$,
starting from a dimensionless radius $x_0$ (note that for the envelope the maximum 
ejecta radius differs from that of the core, thus, $x_0$ is different in the core and 
in the envelope). In such an ejecta the spatial part of the temperature profile was
derived self-consistently by \citet{Blinnikov93} (BP93) and it can be approximately given as
\begin{eqnarray}
\psi(x) &=& \frac{\sin (\alpha x/x_0 )}{\alpha x/x_0}\qquad  (x < x_0) \\
\psi(x) &=& \frac{\sin ( \alpha)}{\alpha} \left(\frac{x}{x_0}\right)^{-(n+1)}\qquad  (x > x_0)
\end{eqnarray}
where the eigenvalue $\alpha$ depends on the power-law exponent as 
$\tan(\alpha) ~\approx~ -\alpha / n$. More details and a mathematically ri\-go\-rous
description can be found in \citet{Blinnikov93} (BP93). For $n=2$, we get $\alpha \approx 2.29$,
which differs from the ``radiative zero'' value of $\alpha = \pi$.
Thus, in the envelope, the temperature in the $x < x_0$ region is similar but
not exactly the same as the simple Arnett-solution given above. This is even more
pronounced above $x_0$, where the BP93 profile is approximately a
power-law with the index of $n+1$. The left panel in Fig.~\ref{fig:temp} shows
the comparison of the Arnett- and the BP93 $\psi(x)$ profiles for
the $n=2$ power-law atmosphere. It is seen that the latter function decreases
outward faster, but it does not converge to 0 at $x=1$ (i.e. at maximum ejecta radius).

However, despite these differences between the tempe\-ra\-ture profiles in the
envelope, its effect on the final light curve is small. Within our modeling scheme 
it affects only the outer envelope, which is thought to contain much less mass and 
has much lower initial density than the core component. Thus, for numerical simplicity, we
decided to apply the simple Arnett-profile for both the core and the envelope,
but note that the estimated parameters of the envelope are somewhat affected
by the choice of the temperature profile, and should be treated with caution. 
Nevertheless, since the envelope parameters are quite weakly constrained (see below) 
this should not be a major concern.

The right panel of Fig.~\ref{fig:temp} exhibits the full normalized temperature profile
($T(r) \sim \psi(r)^{1/4}$) as a function of ejecta radius, after joining the core and 
the envelope components together.
The boundary between the core and the envelope is indicated by the dashed vertical line.
The left-hand side of this panel shows a typical Type IIb configuration (using SN~1993J as a reference), while the right-hand side displays 
a Type IIP model (based on SN~2013ej). It is seen that joining the two components creates
a rather artificial temperature distribution having an abrupt jump at the interface
bet\-ween the core and the envelope. However, the whole configuration looks more-or-less
similar to a temperature profile of an ejecta having an extended envelope with quite
steeply decreasing density profile, as illustrated by the red dashed line 
corresponding to a Blinnikov-Popov temperature profile having $n=7$. Thus, our
two-component configuration, although with an approximate and simplistic temperature
distribution, might mimic more-or-less the expected temperature profiles of supergiant
stars having shallow inner and rapidly decreasing outer density profiles.

Moreover, according to \cite{Popov} the simple Arnett-profile can be used only if the total optical depth of the envelope ($\tau_* = 1/4 \beta s$) at a given time is higher than 1, i.e.  $\tau_* > 1.0$. In this case $\beta = v_{exp}/c$ and $s = (t+t_h)^2/t_d^2$, where $t_h$ and $t_d$ is the hydrodynamic and the effective diffusion time-scale, respectively \citep{Arnett80,Popov}. In our models the validity of this criterion is checked at the end of the plateau phase, when the effect of recombination becomes negligible. Using the parameters derived for the modeled SNe (see below), for the Type IIP SNe $\tau_* > 4$ have been found, while $\tau_* \sim 2$ have been revealed for the Type IIb mo\-dels. Thus, our Type IIP models fully, the Type IIb models marginally satisfy the condition for the photon diffusion approximation.   

\section{Effects of the constant opacity approximation}
\subsection{Calculating the average opacity from SNEC}
One of the strongest simplification in semi-analytic LC models is the assumption of the constant Thompson-scattering opacity ($\kappa$), which can be defined as the average opacity of the ejecta. In this subsection we approximate the average opacity via synthesized Type IIP and Type IIb light curve models. 

We adopt a model star that has a mass of $\sim 18.5$ $\mathrm{M_{\odot}}$ at core collapse. The internal structure of this star was derived from a 20 $\mathrm{M_{\odot}}$ zero-age main-sequence star using the 1D stellar evolution code MESA \citep{Paxton}. The MESA model assumes a non-rotating, non-magnetic stellar configuration with solar metallicity and significant ($10^{-4}\ \mathrm{M_{\odot}/yr}$) mass-loss. The 'Dutch' wind scheme for massive stars was used to model the mass-loss during the AGB and RGB phase, which scenario combines the results form \cite{Glebbeek}, \cite{Vink} and \cite{Nugis}. The opacity calculation in MESA is based on the combination of opacity tables from OPAL,  \cite{Ferguson}, and \cite{Cassisi}.

The evolution of the model star was calculated by MESA until core collapse, and this ori\-ginal model was used for further calculations in producing a Type IIP SN light curve. Moreover, we also built a second model for studying the light curve of Type IIb SN. In order to estimate the progenitor of a Type IIb SN, most of the outer H-rich envelope of the original MESA model was removed ma\-nually, so that only $\sim 1$ M$_{\odot}$ envelope mass remained. The subsequent hydrodynamic evolutions were followed by SNEC, which is a 1D Lagrangian supernova explosion code \citep{Morozova}. SNEC solves the hydrodynamics and diffusion radiation transport in the expanding envelopes of CCSNe, taking into account recombination effects and the decay of radioactive nickel. During the calculations the ``thermal bomb'' explosion scheme was used, in which the total energy of the explosion is injected into the model with an exponential decline both in time and mass coordinate. The SNEC code calculates the opacity in each grid point of the model from Rosseland mean opacity tables for different chemical compositions, temperatures and densities. During this process an opacity minimum was also taken into account by the code. In our simulations this opa\-city edge was $0.24$ and $0.01$ $\mathrm{cm^2/g}$ for the pure metal and the solar composition envelope \citep{Bersten11}, respectively. Thus, in SNEC the opacity at each time and grid point is chosen as the maximum value between the calculated Rosseland mean opacity and the opacity edge for the corresponding composition.

\begin{figure}[!ht]
\includegraphics[width=9cm]{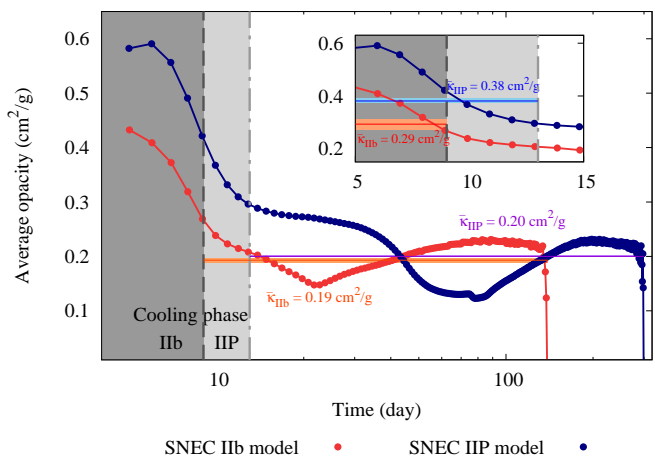}
\caption{The dependence of $\kappa(M_{ph})$ on the time for Type IIP (blue) and Type IIb (red) SNEC model.}
\label{fig:kappa}
\end{figure}

To estimate the average opacity for Type IIP and Type IIb SNe the original SNEC opacity output file was used. At a given time we defined $\kappa(M_{ph})$ by integrating the opacity from the mass coordinate of the neutron star ($M_0 = 1.34 M_\odot$) up to the mass coordinate of the photosphere ($M_{ph}$) as
\begin{equation}
\kappa(M_{ph}) = {\frac{1}{M_{ph} - M_0}} \int\limits_{M_0}^{M_{ph}} \kappa\ dm\ .
\end{equation}
Taking into account that our semi-analytic model uses the same opacity when calculating the entire light curve, we defined the average opacity ($\overline{\kappa}$) by integrating $\kappa(M_{ph})$ from several day after the shock breakout ($t_0$) up to $t_{end}$ as
\begin{equation}
\overline{\kappa} = {\frac{1}{t_{end} - t_0}} \int\limits_{t_0}^{t_{end}} \kappa(M_{ph})\ dt\ .
\end{equation}
Fig.~\ref{fig:kappa}. shows the dependence of $\kappa(M_{ph})$ on the time. To receive comparable results with our two-component estimates, we separately calculate the average opacities for both the early cooling phase and the photospheric phase. In the cooling phase $t_0 = 5$ days, while $t_{end}$ was chosen as the appro\-ximate termination of the cooling phase, when the opacity drops rapidly, which was 9 days and 13 days for Type IIb and Type IIP model, respectively. For the photospheric phase $t_0$ was defined to be equal to $t_{end}$ of the cooling phase, and we integrate up to the end of the nebular phase.  Vertical gray lines in Fig.~\ref{fig:kappa}. represent these time boundaries for both Type IIb and Type IIP SNe. Horizontal lines indicate the different $\overline{\kappa}$ values as the average opacities of different phases and models. It can be seen that in the cooling phase $\overline{\kappa}$ is $\sim 0.4\ \mathrm{cm^2/g}$ for a Type IIP SN with a massive H-rich ejecta. However, the average opacity decreases to $\sim 0.3\ \mathrm{cm^2/g}$ for a Type IIb, which corresponds to a star that lost most of its H-rich envelope. In contrast, during the later phase the average opacity of Type IIP and Type IIb is considerably similar, having a value of $\sim 0.2\ \mathrm{cm^2/g}$.  

Note that the constant Thompson-scattering opacity is not an adequate approximation either at early or at late phases because of the rapidly changing opacities, but the calculated average opacities show a reasonably good agreement with previous studies \citep[e.g.,][]{Nakar, Huang}, where $\overline{\kappa} = 0.24$ $\mathrm{cm^2/g}$ was used for modeling the LCs of Type IIP SNe. Thus, in the following analysis we use the average opacities ($\overline{\kappa} = 0.3$ $\mathrm{cm^2/g}$ and $\overline{\kappa} \sim 0.4$ $\mathrm{cm^2/g}$) from SNEC to approximate $\kappa$ in the shell models of Type IIb and Type IIP SNe, respectively. To estimate $\kappa$ for the core model we take into account the derived expansions velocities and the $\overline{\kappa}$ values form SNEC. Because the range of the varying $\kappa$ values is narrower for the Type IIb than for the Type IIP configuration (see Fig.~\ref{fig:kappa}), then for Type IIb SNe we use $\kappa = 0.2$  $\mathrm{cm^2/g}$, while for Type IIP SNe $\kappa \approx 0.2 \pm 0.1 $  $\mathrm{cm^2/g}$ was chosen.

\subsection{Correlation between the opacity and ejected mass}
In this subsection we examine the effect of the constant opacity approximation, which may have an important role in analysing the fitting results, via a synthesized Type IIP light curve. 

In this case we use the Type IIP SNEC model discussed above (Section 3.1). The SNEC model light curve is compared with the light curves calculated by the semi-analytic code. Two models were computed with the latter: in Model A we used exactly the same physical parameters as in the SNEC model, but varied $\kappa$ to get the best match between the two LCs. In Model B we applied $\kappa = 0.2$ $\mathrm{cm^2/g}$, similar to the average opacity of the SNEC model, but tweaked the other parameters until reasonable match between the data and the model LC was found.

\begin{figure}[!ht]
\includegraphics[width=9cm]{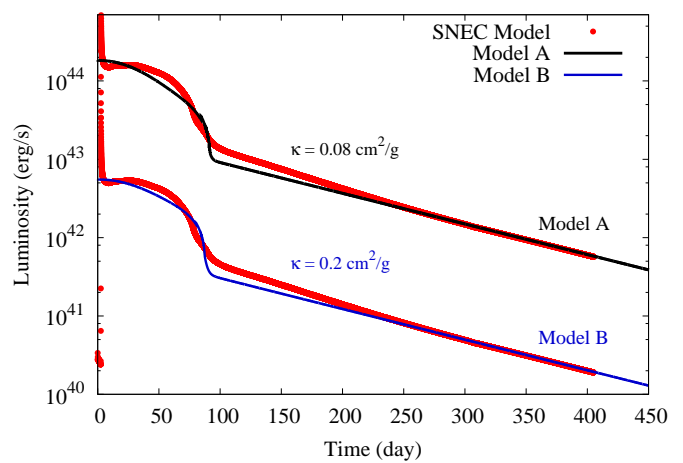}
\caption{Comparison of the bolometric light curve from SNEC (red) with the semi-analytic LCs. Model A (black) and Model B (blue) represent the best LC fit with low and high opacity, respectively (Table~\ref{table:opacity}). The LC of Model A was shifted vertically for better visibility.}
\label{fig:snec}
\end{figure}

The parameters of the hydrodynamic model and from the semi-analytic LC models are summarized in Table~\ref{table:opacity}. Even though the opacities are really different in the two models, both light curves show acceptable match with the SNEC light curve (Fig.~\ref{fig:snec}). These results represent the well-known issue that the analytic codes having constant Thompson-scattering opacity usually predict lower ejected masses than hydrodynamic calculations \citep[e.g.,][]{Utrobin09, Smartt}. This issue is probably due to the incorrect assumption of constant opacity as well as the reduced dimension in the hydrodynamic simulations. On the contrary, to get the same ejected mass in both the hydrodynamic and analytic calculations, an extremely low Thompson-scattering opacity is needed. This low opa\-city value may be a hint of a dominance of metals over H, which looks difficult to explain in the outer region of the H-rich ejecta. 

\begin{table}[!h]
\caption{Model parameters for the synthetic LCs} 
\label{table:opacity}     
\centering                  
\begin{tabular}{l c c c }          
\hline
\hline \\                      
Parameter &  SNEC & Model A & Model B \\
\hline \\                        
$R_0$ ($10^{13}$ cm)& 7.66 & 7.66 & 7.90  \\
$M_{ej}$ ($\mathrm{M_\odot}$)& 14.0 & 14.0 & 8.9  \\
$M_{Ni}$ ($\mathrm{M_\odot}$)& 0.05 & 0.05 &  0.05 \\
$E_{tot}$ ($10^{51}$ erg)& 2.0 & 2.0 & 2.3 \\
$\kappa$ ($\mathrm{cm^2/g}$)& 0.2 & 0.08  & 0.2 \\
\hline 
\hline                                            
\end{tabular}
\end{table}

Furthermore, to explain the possible reason for the previously mentioned ejected mass problem, we suspect that this effect is due to parameter correlation \citep{Nagy}. As it can be seen in Table~\ref{table:opacity} only two parameters change drastically in the models, which suggests that the correlation bet\-ween $\kappa$ and $M_{ej}$ may have a major role in this problem. 

\begin{figure}[!ht]
\includegraphics[width=9cm]{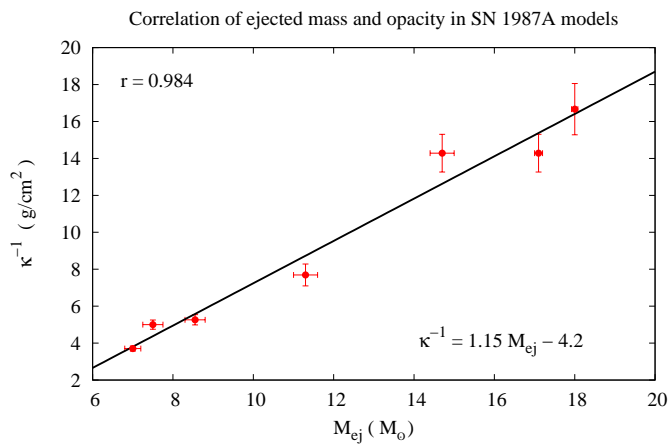}
\caption{Correlation between the opacity and the ejected mass for SN 1987A models.}
\label{fig:corr}
\end{figure}

To examine the effect of this correlation, we used the available hydrodynamic and analytic model parameters ($R_0,\ M_{ej},\ E_{kin}$) of SN 1987A (see the details in Sec. 6.). SN 1987A was chosen, because this object is the best studied SN ever, thus, several different model calculations are available in the li\-terature. To estimate the opacity of these different models, we took the parameters from each published SN 1987A model, fixed them "as-is", and fit the LC with our code while tweaking only the values of $\kappa$. The final results can be seen in Fig. ~\ref{fig:corr}, where each red dot represents one of the applied models. It is seen that $M_{ej}$ and $\kappa^{-1}$ are strongly correlated parameters.

To illustrate a quantitative measure of the correlation we calculate the correlation coefficient as
\begin{equation}
r = \frac{\sum\limits_{i=1}^{n}\ (M_{i} - \overline{M})\ (\kappa^{-1}_i - \overline{\kappa}\ ^{-1})}{(n - 1)\ \sigma_M\ \sigma_\kappa} = 0.984\ ,
\end{equation}
where $\sigma_M$, $\sigma_\kappa$ and $\overline{M}$, $\overline{\kappa}\ ^{-1}$ are the standard deviations and the mean values of $M_{ej}$ and $\kappa^{-1}$, res\-pec\-ti\-vely. Because $r$ is close to 1 the ejected mass and opacity are highly correlated, which cause significant uncertainty in the determination of these parameters. Clearly, neither $M_{ej}$ nor $\kappa$ can be reliably inferred from these LC models, only their product, $M_{ej}\cdot \kappa$, is constrained \citep[see also][]{Wheeler} 

\section{Effect of the uncertainty of the explosion time on the estimated parameters}
When pre-explosion observations of the SN site are not available, the explosion time of the supernova can be rather uncertain. Determining the moment of first light of a supernova explosion is very important, because the shape of the rising part of the light curve depends critically on the exact date of the explosion. The physical parameters inferred from such a light curve may suffer from large systematic errors.
\begin{table*}[!ht]
\caption{The effect of the uncertainty of the explosion time on the fit parameters for a Type IIb SN (based on the LC of SN 1993J)} 
\label{table:exp1}     
\centering                  
\begin{tabular}{l c c c c c c c}          
\hline
\hline \\                      
Explosion time  & $t_0 + 4d$ & $t_0 + 3d$ & $t_0 + 2d$ & $t_0 + 1d$  & $t_0$ &  $t_0 - 1d$ &  $t_0 - 2d$ \\
\hline \\                    
$R_\mathrm{core}$ ($10^{11}$ cm)& 2.1 & 2.0 & 1.9 & 1.8 & 1.7 & 1.6 & 1.5\\
$M_\mathrm{core}$ ($\mathrm{M_\odot}$)& 1.20 & 1.25 & 1.30 & 1.35 &  1.40 &  1.45 & 1.50\\
$E_\mathrm{core}$ ($10^{51}$ erg)& 4.2 & 4.2 & 4.2 & 4.2 & 4.2 & 4.2 & 4.2 \\
$M_\mathrm{Ni}$ ($\mathrm{M_\odot}$)& 0.094 & 0.096 & 0.098 & 0.1 &  0.1 &  0.1 & 0.1\\
$R_\mathrm{shell}$ ($10^{13}$ cm)& 3.0 & 3.0 & 3.0 & 3.0 & 3.0 & 3.0 & 3.0\\
$M_\mathrm{shell}$ ($\mathrm{M_\odot}$)& 0.06 & 0.07 & 0.08 & 0.09 &  0.10 &  0.11 & 0.12\\
$E_\mathrm{shell}$ ($10^{51}$ erg)& 0.74 & 0.75 & 0.76 & 0.78 & 0.80 & 0.81 & 0.83 \\
\hline
\hline
\end{tabular}                                            
\end{table*} 

\begin{figure}[!ht]
\includegraphics[width=9cm]{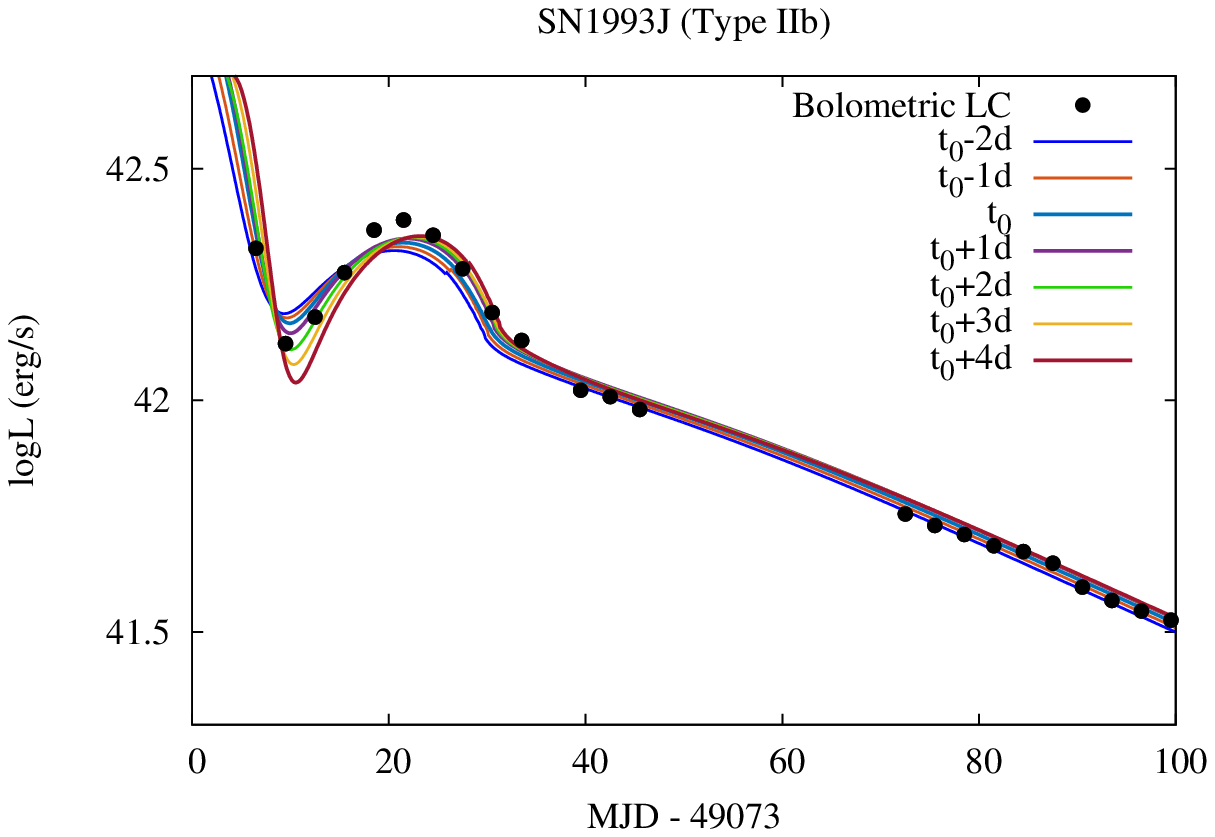}
\caption{Comparison of the bolometric light curve of the Type IIb SN 1993J (dots) with model LCs calculated with different explosion dates (lines).}
\label{fig:exp93J}
\end{figure}

\begin{figure}[!ht]
\includegraphics[width=9cm]{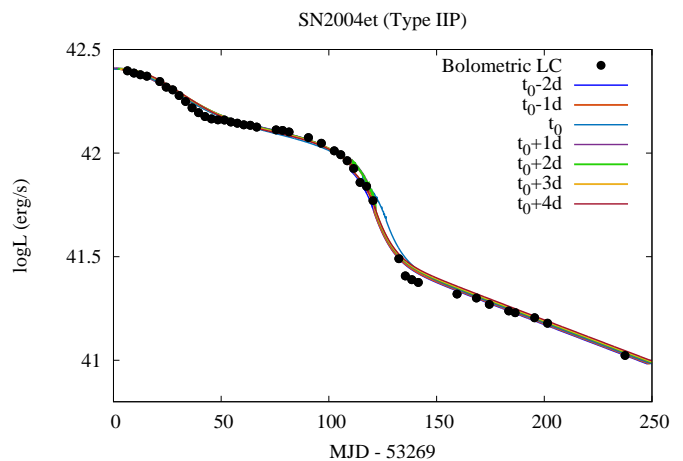}
\caption{Comparison of the bolometric light curve of the Type IIP SN 2004et (dots) with model LCs calculated with different explosion dates (lines).}
\label{fig:exp}
\end{figure}

This effect can be even more significant in the case of SNe with extended envelopes. 
In order to estimate the uncertainty of model parameters due to the unknown date of explosion, we fit the bolometric light curve of a Type IIb (Fig.~\ref{fig:exp93J}) and a Type IIP (Fig.~\ref{fig:exp}) SN assuming different explosion dates. For these calculations we use the two-component configuration described in Section 2. 

It is apparent from Table~\ref{table:exp1} and ~\ref{table:exp2} that for Type IIP SNe an uncertainty of about 7 days in the explosion date generates mo\-derate (5 - 10 $\%$) relative errors in the derived masses of the inner core ($M_{\mathrm{core}}$) and the outer envelope ($M_{\mathrm{shell}}$), and the initial radius of the core ($R_\mathrm{core}$). For Type IIb SNe only $M_\mathrm{Ni}$ and the total energy of the outer envelope ($E_\mathrm{shell}$) show similar uncertainties. In this case the uncertainty of the derived $M_{\mathrm{shell}}$, $R_\mathrm{core}$ and $M_\mathrm{core}$ may increase up to about 50$\%$, 40$\%$ and 20$\%$, respectively.

\begin{table*}[!ht]
\caption{The effect of the uncertainty of the explosion time on the fit parameters for a Type IIP SN (based on the LC of SN 2004et)} 
\label{table:exp2}     
\centering                  
\begin{tabular}{l c c c c c c c}          
\hline
\hline \\                      
Explosion time  & $t_0 + 4d$ & $t_0 + 3d$ & $t_0 + 2d$ &  $t_0 + 1d$ &  $t_0$ & $t_0 - 1d$ & $t_0 - 2d$ \\
\hline  \\                         
$R_\mathrm{core}$ ($10^{13}$ cm) & 3.8 & 3.9 & 4.0 & 4.1 & 4.1 & 4.2 & 4.3 \\
$M_\mathrm{core}$ ($\mathrm{M_\odot}$)& 10.7 & 10.8 &  10.9 &  11.0 & 11.1 & 11.2 & 11.3 \\
$E_\mathrm{core}$ ($10^{51}$ erg)& 1.95 & 1.95 & 1.95 & 1.95 & 1.95 & 1.95 & 1.95 \\
$M_\mathrm{Ni}$ ($\mathrm{M_\odot}$)& 0.06 & 0.06 & 0.06 & 0.06 &  0.06 &  0.06 & 0.06\\
$R_\mathrm{shell}$ ($10^{13}$ cm)& 6.8 & 6.8 & 6.8 & 6.8 & 6.8 & 6.8 & 6.8\\
$M_\mathrm{shell}$ ($\mathrm{M_\odot}$)& 1.06 & 1.07 &  1.08 &  1.09 & 1.1 & 1.11 & 1.12 \\
$E_\mathrm{shell}$ ($10^{51}$ erg)& 1.29 & 1.29 & 1.29 & 1.29 & 1.29 & 1.29 & 1.29 \\
\hline
\hline
\end{tabular}                                            
\end{table*}  

\section{Model fits to Type IIb supernovae}
In the following sections we apply the two-component LC model for real SNe, both Type IIb and IIP.
We use the bolometric light curves of the observed SNe, assembled in a similar way as in our 
previous paper \citep{Nagy}, and compare them with model light curves computed from the two-component
model detailed above. Note that formal $\chi^2$-fitting was not performed, as the strong correlation
between the parameters would make such a fit ill-constrained \citep[see e.g.][]{Nagy}. 
Instead, the model parameters were varied manually until reasonable agreement between the data 
and the model was found.

To estimate the physical properties of Type IIb SNe, we fit the light curves of two well observed events, SN 1993J and SN 2011fu. SN 1993J was discovered in NGC 3031 (M81) on 1993 March 28.9 UT by F. Garcia \citep{Ripero}. UBVRI photometry was presented by \cite{Richmond}, who determined the moment of
explosion as JD 2449073.
SN 2011fu was discovered in a spiral galaxy UGC 1626 by F. Ciabattari and E. Mazzoni \citep{Ciabattari} on 2011 September 21.04 UT. The estimated explosion date is 2011 September 18.0 (JD 2455822.5).

To reach better agreement with spectroscopic observations, a constant density profile was chosen in the inner core, while we use a power-law density structure ($n=2$) in the outer envelope. 
The inner boundary ($x_0$) of the density profile was set as $x_0^{core} = 0.1$ and $x_0^{shell} = 0.4$. We use $\kappa = 0.2$  $\mathrm{cm^2/g}$ for the inner, H-poor core and $\kappa = 0.3$  $\mathrm{cm^2/g}$ for the H-He outer envelope, which are consistent with the estimated average opacities from SNEC Type IIb model. 

\begin{figure}[!ht]
\includegraphics[width=9cm]{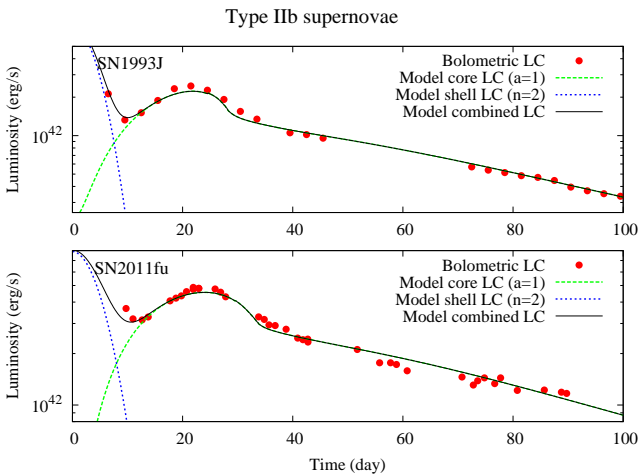}
\caption{Comparison of the bolometric light curves of Type IIb SNe (dots) with the best two-component model fits. The green and blue curves represent the contribution from He-rich core and the extended H-envelope, respectively, while the black lines show the combined LCs. }
\label{fig:IIb}
\end{figure}

Due to their lower ejected masses, the gamma-ray leakage may be significant in Type IIb SNe. Typical $A_g$ values for these events are between $10^{3}$ and $10^4$ $\mathrm{day^2}$. The other parameters of the best fit-by-eye models are summarized in Table~\ref{table:IIb}, and the LCs are plotted in Fig.~\ref{fig:IIb}. From these results it can be seen that the mass of the He-rich core is $\sim 1$ $\mathrm{M_\odot}$, while the extended envelope contains $\sim 0.1$ $\mathrm{M_\odot}$, although the latter depends on the chosen value of $x_0$ in the envelope, which is poorly constrained. 
Using $x_0 \sim 0.1$ instead of $0.4$ in the envelope would require somewhat higher energies 
and an order of magnitude less mass ($\sim 0.01\ \mathrm{M_\odot}$) to fit the initial peak. 

\begin{table}[!h]
\caption{Model parameters of Type IIb SNe} 
\label{table:IIb}     
\centering                  
\begin{tabular}{l c c c c }          
\hline
\hline \\                      
Parameter &  \multicolumn{2}{c}{SN 1993J} & \multicolumn{2}{c}{SN 2011fu}\\ 
 & core & envelope & core & envelope \\
\hline \\                        
$R_0$ ($10^{12}$ cm)& 0.35 & 30 & 0.35 & 13 \\
$M_{ej}$ ($\mathrm{M_\odot}$)& 2.15 & 0.1 & 2.2 & 0.12 \\
$M_{Ni}$ ($\mathrm{M_\odot}$)& 0.1 & - & 0.23 &  - \\
$E_{tot}$ ($10^{51}$ erg)& 3.7 & 0.8 & 3.4 & 0.8 \\
$E_{kin}/E_{th}$ & 1.85 & 7.0 & 2.4 & 1.67 \\
$\kappa$ ($\mathrm{cm^2/g}$)& 0.2 & 0.3 & 0.2 & 0.3 \\
\hline 
\hline                                            
\end{tabular}
\end{table}

\subsection{Comparison with other models}
For the examined Type IIb SNe the physical parameters of both the envelope and the core were determined by se\-ve\-ral authors. Here we compare all of the available parameters with the values determined by the two-component LC fitting, keeping the same terminology for the parameters from the two-component models that was used in Section 4. The only exception is that $E_{core}$ refers to the kinetic energy of the core component.

The physical properties of SN 1993J were calculated by se\-veral LC modeling codes \citep{ Shigeyama94, Utrobin94, Woosley, Young, Blinnikov98}. First, the physical properties of this explosion was determined by a hydrodynamic calculation of \cite{Shigeyama94} as $R_{shell} = (1.7 - 2.5)\cdot 10^{13}$ cm, $E_{core} = 1.0 - 1.2$ foe, while the mass of the extended envelope is below $\sim 0.9$ $\mathrm{M_\odot}$. The fundamental parameters of SN 1993J were also inferred by \cite{Utrobin94}, who found that the radius of the progenitor was $\sim 3.2 \cdot 10^{12}$ cm, the ejected mass was $\sim$ 2.4 $\mathrm{M_\odot}$, the nickel mass was $\sim$ 0.06 $\mathrm{M_\odot}$, and the energy of the explosion was $\sim$ 1.6 foe.

Another scenario was modeled by the KEPLER stellar evolution and hydrodynamic code based on the assumption that the progenitor of SN 1993J lost its outer H envelope due to mass transfer to a binary companion \citep{Woosley}. At the time of the explosion the H envelope mass was $ 0.2 \pm 0.05 \mathrm{M_\odot}$ and the radius of the star was ($4 \pm 1)\cdot 10^{13}$ cm. In this model $0.07 \pm 0.01 \mathrm{M_\odot}$ of $^{56}$Ni was produced during the explosion, and the ejected mass was found to be about 1.2 $\mathrm{M_\odot}$.

The radius of the progenitor star was estimated by \cite{Young} as $R_{shell} = (2 - 4)\cdot 10^{13}$ cm, while the nickel mass was found to be $M_{Ni} \sim$ 0.1 $\mathrm{M_\odot}$. The ejected mass settled in the range of 1.9 - 3.5 $\mathrm{M_\odot}$. \cite{Young} used a H-rich atmosphere with $M_{shell} \sim$ 0.1 - 0.5 $\mathrm{M_\odot}$ and $R_{shell} \sim$ $10^{13}$ cm to represent the extended envelope due to the mass-loss of the progenitor just before the explosion.

The explosion of SN 1993J was also calculated with both STELLA and EDDINGTON hydrodynamic codes \citep{Blinnikov98}. 
In these models $0.073 \mathrm{M_\odot}$ of $^{56}$Ni was found, and $1.55 \mathrm{M_\odot}$ mass was ejected with $E_{core} \sim 1.2$ foe. Although these four calculations applied different scenarios, the obtained results are in the same parameter range and also show a good agreement with our model parameters, except for the kinetic energy (Table~\ref{table:93J}) which is usually overestimated by the semi-analytic LC fitting codes. 
 
\begin{table*}[!ht]
\caption{Model parameters for SN 1993J} 
\label{table:93J} 
\centering                  
\begin{tabular}{l c c c c c c}          
\hline
\hline \\                      
Parameter &\multicolumn{5}{c}{Literature} & This paper\\
& S94\tablefootmark{1} & U94\tablefootmark{2} & W94\tablefootmark{3} & Y95\tablefootmark{4} & B98\tablefootmark{5} &\\ 
\hline  \\                         
$R_{shell}$ ($10^{13}$ cm)& 2.1 & 3.2 & 4.0 & 3.0 & 4.33 & 3.0\\
$M_{shell}$ ($\mathrm{M_\odot}$)& < 0.9 & 0.1 & 0.2 & 0.3 & 0.2 & 0.1\\
$M_{core}$ ($\mathrm{M_\odot}$)& 4.0 & 2.4 & 1.2 & 2.7 & 2.3 & 2.15\\
$M_{Ni}$ ($\mathrm{M_\odot}$)& 0.08 & 0.06 & 0.07 & 0.1 & 0.073 & 0.1\\
$E_{core}$ ($10^{51}$erg)& 1.1 & 1.6 & 1.3 & 1.0 & 1.2 & 2.4\\
\hline 
\hline                                            
\end{tabular}
\tablebib{(1)~\citet{Shigeyama94}; (2)~\citet{Utrobin94}, (3)~\citet{Woosley}, (4)~\citet{Young}, (5)~\citet{Blinnikov98}.}
\end{table*}

A hydrodynamic calculation for SN 2011fu was presented by \cite{Morales}, assuming $R_{shell} = 3.13\cdot 10^{13}$ cm,  $M_{core} = 3.5$ $\mathrm{M_\odot}$, $E_{core} = 1.3$ foe and $M_{Ni} = 0.15$ $\mathrm{M_\odot}$. The significant differences between these values and our estimates are probably due to using different distances and extinctions during the calculation of the bolometric light curve. Nonetheless, our approximate parameters are in the same order-of-magnitude as the hydrodynamic results (Table~\ref{table:11fu}).

Note that the light curve of SN 2011fu was also fit by \cite{Kumar} with the analytic model of \cite{Arnett}. They derived $R_{core} = 2\cdot 10^{11}$ cm, $M_{core} = 1.1$ $\mathrm{M_\odot}$, $M_{Ni} = 0.21$ $\mathrm{M_\odot}$ and $E_{core} = 2.4$ foe for the inner He-core, while for the outer hydrogen envelope $R_{shell} = 10^{13}$ cm, $M_{shell} = 0.1$ $\mathrm{M_\odot}$ and $E_{shell} = 0.25$ foe were found. These estimated values are in a good agreement with our results, which is expected because both models apply similar physical modeling schemes. The minor differences in the envelope parameters are due to the differences between the adopted density profiles, because \cite{Kumar} use an exponential profile (a = 1) against our constant density model. Note that we also tested the application of the exponential density profile, but the shape of the generated LC showed better agreement with the observed data in the constant density model, which is in accord with the results of \cite{Arnett}.    

\begin{table}[!ht]
\caption{Model parameters for SN 2011fu} 
\label{table:11fu} 
\centering                  
\begin{tabular}{l c c c }          
\hline
\hline \\                      
Parameter &\multicolumn{2}{c}{Literature} & This paper\\
& K13\tablefootmark{1} & MG15\tablefootmark{2} & \\ 
\hline  \\                         
$R_{shell}$ ($10^{13}$ cm)& 1.0 & 3.13 & 1.3\\
$M_{shell}$ ($\mathrm{M_\odot}$)& 0.1 & 0.3 & 0.12\\
$M_{core}$ ($\mathrm{M_\odot}$)& 1.1 & 3.5 & 2.2\\
$M_{Ni}$ ($\mathrm{M_\odot}$)& 0.21  & 0.15 & 0.23\\
$E_{core}$ ($10^{51}$ erg) & 2.4 & 1.3 & 2.4 \\
\hline 
\hline                                            
\end{tabular}
\tablebib{(1)~\citet{Kumar}; (2)~\citet{Morales}.}
\end{table}

\section{Model fits to Type IIP supernovae}

In order to derive the model parameters of Type IIP SN, we fit the LCs of SNe 1987A, 2003hn, 2004et, 2005cs, 2009N, 2012A, 2012aw, and 2013ej. Table~\ref{table:host} shows the observational properties of these events.

 \begin{table*}[!ht]
\caption{The observational properties of Type IIP SNe} 
\label{table:host}     
\centering                  
\begin{tabular}{l l c c c }          
\hline
\hline \\                      
SN & Discovery date (UT) & Host & MJD of explosion & Reference \\ 
\hline \\                        
1987A & 1987 Feb. 24.23 & LMC & 46850.55 & 1, 2\\
2003hn & 2003 Aug. 25.7 & NGC 1448 & 52874.0 & 3, 4\\
2004et & 2004 Sept. 27 & NGC 6946 & 53269.0 & 5\\
2005cs & 2005 June 30 & M51 & 53549.0 & 6\\
2009N & 2009 Jan. 24.86 &  NGC 4487 & 54848.1 & 7\\
2012A & 2012 Jan. 7.39 & NGC 3239 & 55929.0 & 8\\
2012aw & 2012 Mar. 16.86 & M95 & 56002.5 & 9\\ 
2013ej & 2013 July 25.45 & M74 & 56497.3 & 10, 11\\ 
\hline 
\hline                                            
\end{tabular}
\tablebib{(1)~\citet{Kunkel}, (2)~\citet{Suntzeff}, (3)~\citet{Evans}, (4)~\citet{Krisciunas}, (5)~\citet{Zwitter}, (6)~\citet{Kloehr},(7)~\citet{Nakano}, (8)~\citet{Moore}, (9)~\citet{Fagotti}, (10)~\citet{Kim}, (11)~\citet{Dhungana}.}
\end{table*}

While fitting these SNe, we use a constant density profile for the inner core, while the outer H-rich envelope has a power-law density distribution. The constant-density approximation was found to work surprisingly well for fitting the plateau phase of Type IIP SNe except for the early cooling phase \citep{Arnett, Nagy}. The power-law density profile with $n = 2$  is an acceptable choice if we assume a steady stellar wind, similar to \citet{Moriya}. In the outer shell the energy input from recombination 
was neglected because of the low envelope mass and rapid cooling in this region. However, the effect of the recombination is important in the inner core, because recombination is responsible for the appearance of the entire plateau phase. After the plateau phase, the LC follows the time-dependence of the decay of radioactive cobalt. In most cases gamma-ray lea\-kage was found negligible. The two exceptions are SN 1987A and 2013ej, where the effect of gamma-ray escape was taken into account by setting $A_g \sim$ $2.7 \cdot 10^5$ and $3 \cdot 10^4$ day$^2$, respectively. Note that for SN 1987A the gamma-ray leakage was also examined by \cite{Popov92}. In that study the charac\-te\-ris\-tic time-scale of the gamma-rays ($T_0$) was 500 - 650 days, which shows a reasonably good agreement with our result of $\sqrt{A_g} \approx 520$ days.

\begin{figure}[!ht]
\includegraphics[width=9cm]{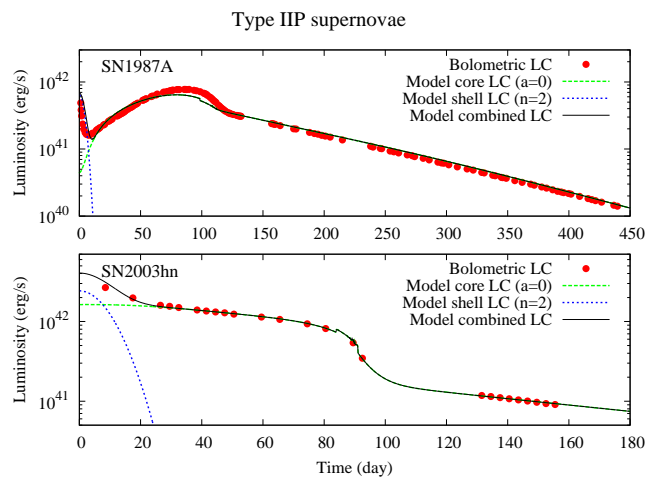}
\caption{Comparison of the bolometric light curves of Type IIP SNe (dots) with the best two-component model fits. The green and blue curves represent the contribution from the inner ejecta and the extended H-envelope, respectively, while the black lines show the combined LCs. }
\label{fig:IIP_1}
\end{figure}

\begin{figure}[!ht]
\includegraphics[width=9cm]{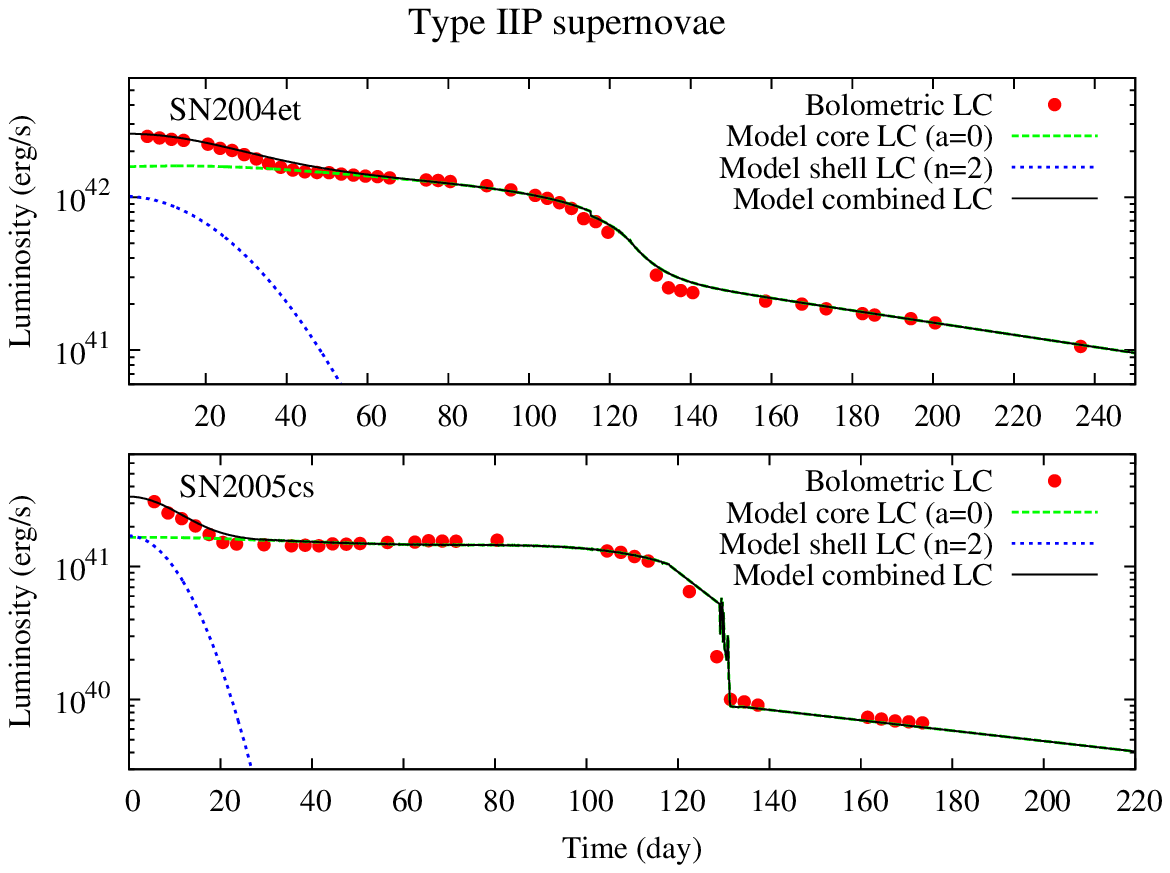}
\caption{Comparison of the bolometric light curves of Type IIP SNe (dots) with the best two-component model fits. The green and blue curves represent the contribution from the inner ejecta and the extended H-envelope, respectively, while the black lines show the combined LCs. }
\label{fig:IIP_2}
\end{figure}

\begin{figure}[!ht]
\includegraphics[width=9cm]{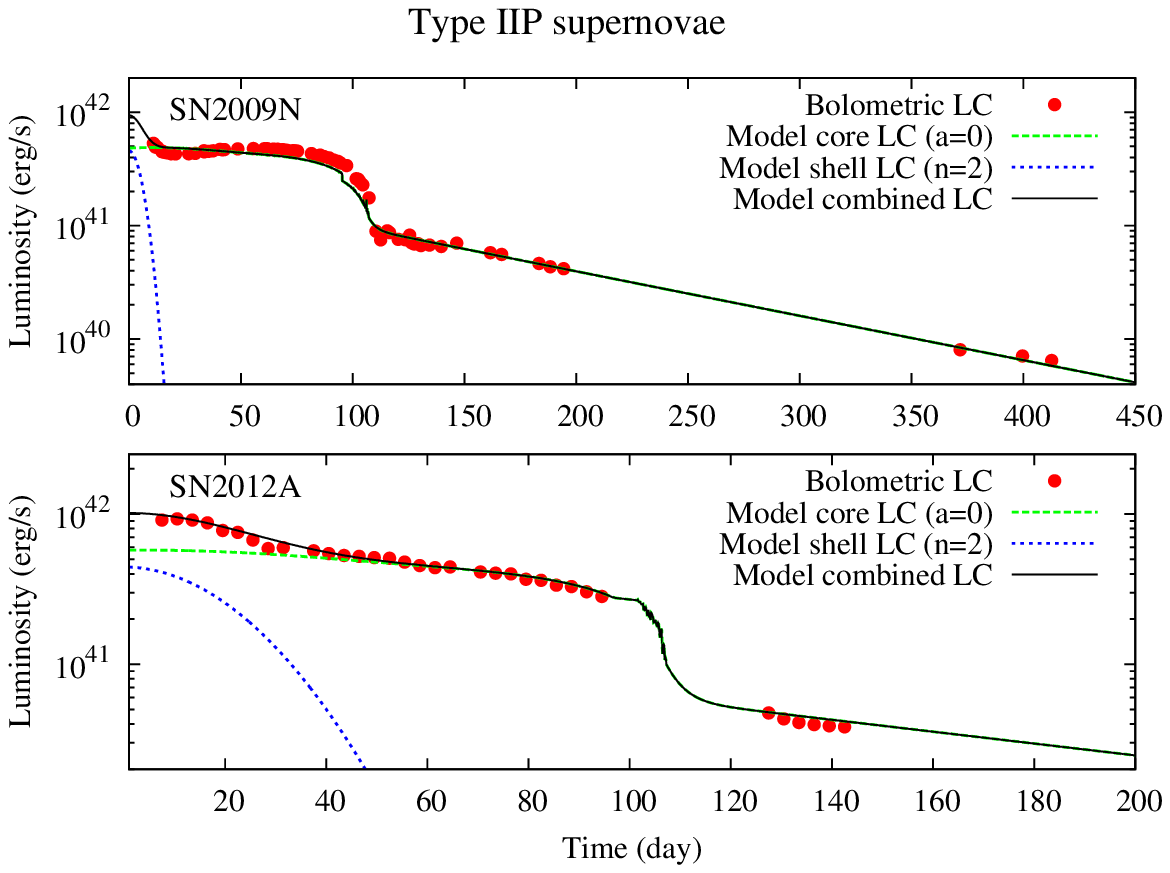}
\caption{Comparison of the bolometric light curves of Type IIP SNe (dots) with the best two-component model fits. The green and blue curves represent the contribution from the inner ejecta and the extended H-envelope, respectively, while the black lines show the combined LCs. }
\label{fig:IIP_3}
\end{figure}

\begin{figure}[!ht]
\includegraphics[width=9cm]{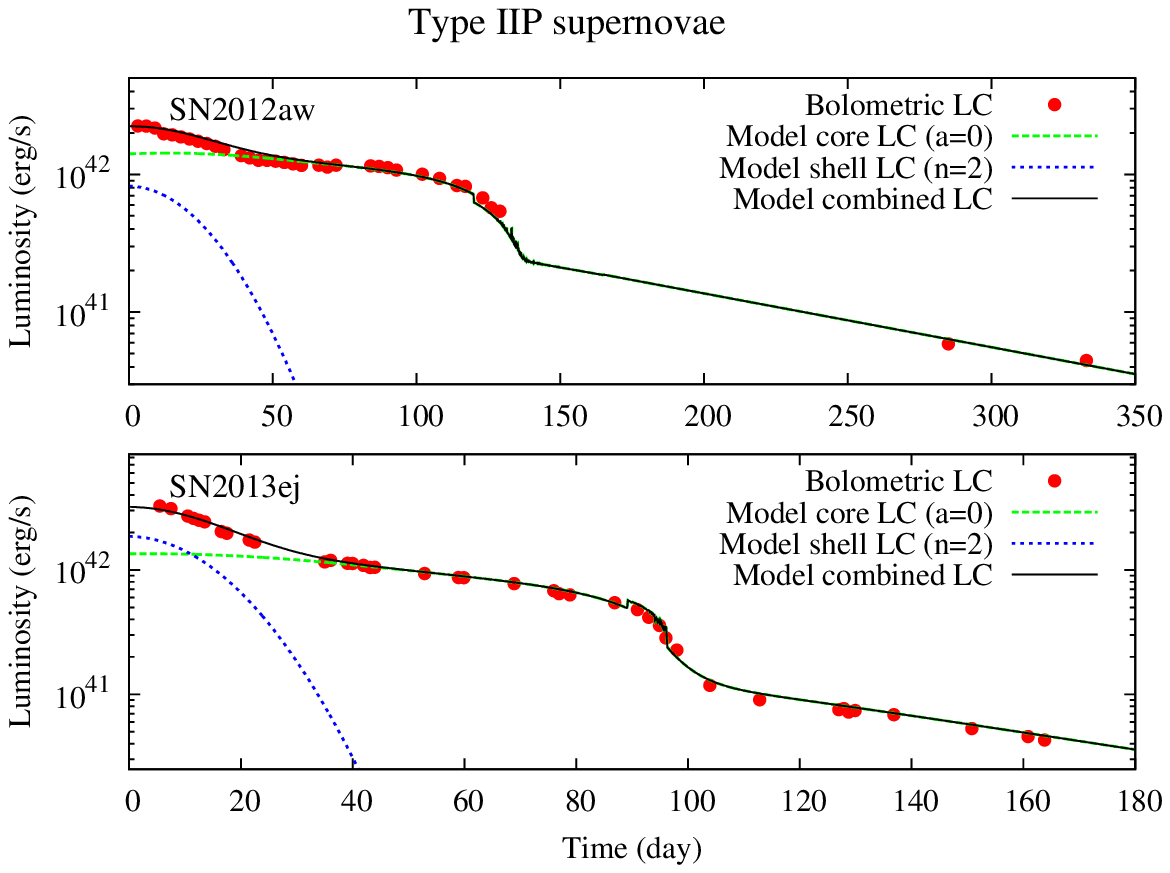}
\caption{Comparison of the bolometric light curves of Type IIP SNe (dots) with the best two-component model fits. The green and blue curves represent the contribution from He-rich core and the extended H-envelope, respectively, while the black lines show the combined LCs. }
\label{fig:IIP_4}
\end{figure}

The best model fits for Type IIP supernovae are summarized in Table~\ref{table:IIP}, and the LCs are plotted in Fig.~\ref{fig:IIP_1} - ~\ref{fig:IIP_4}. It can be seen that the masses of the inner region 
are about 7 - 20 $\mathrm{M_\odot}$, while the masses of the outer envelope are less than 1 $\mathrm{M_\odot}$. The total supernova explosion energies show a huge diversity, but its reality
is questionable as it may be simply due to the strong correlation between the explosion energy, 
the ejected mass and the progenitor radius \citep[see e.g. ][]{Nagy}. 

\begin{table*}[!ht]
\caption{Model parameters of Type IIP SNe} 
\label{table:IIP}     
\centering                  
\begin{tabular}{l c c c c c c c c}          
\hline
\hline \\                      
Parameter &  \multicolumn{2}{c}{SN 1987A} &  \multicolumn{2}{c}{SN 2003hn} & \multicolumn{2}{c}{SN 2004et} & \multicolumn{2}{c}{SN 2005cs}\\ 
& core & envelope & core & envelope & core & envelope & core & envelope\\
\hline  \\                         
$R_0$ ($10^{12}$ cm)& 2.9 & 10 & 16 & 40 & 42 & 68 & 12 & 20\\
$M_{ej}$ ($\mathrm{M_\odot}$)& 8.6 & 0.1 & 10.6 & 0.3 &  11.0 &  1.1 & 8.0 & 0.3\\
$M_{Ni}$ ($\mathrm{M_\odot}$)& 0.069 & - & 0.025 & - &  0.06 &  - & 0.002 & -\\
$E_{tot}$ ($10^{51}$ erg)& 1.52 & 0.42 & 3.2 & 1.2 & 1.95 & 1.29 & 0.48 & 0.814\\
$E_{kin}/E_{th}$ & 11.7 & 20 & 1.67 & 11 & 2.25 & 13.3 & 2.0 & 57.1 \\
$\kappa$ ($\mathrm{cm^2/g}$)& 0.19 & 0.34 & 0.23 & 0.4  & 0.3 & 0.4 & 0.3 & 0.4\\
\hline
\hline\\  
Parameter &  \multicolumn{2}{c}{SN 2009N} & \multicolumn{2}{c}{SN 2012A} & \multicolumn{2}{c}{SN 2012aw} & \multicolumn{2}{c}{SN 2013ej}\\ 
& core & envelope & core & envelope & core & envelope & core & envelope\\
\hline  \\                         
$R_0$ ($10^{12}$ cm)& 14 & 30 & 17 & 40 & 29.5 & 45 & 29 & 68 \\
$M_{ej}$ ($\mathrm{M_\odot}$)& 7.5 & 0.12 & 8.0 & 0.82 & 20 & 1.0 & 10 & 0.6 \\
$M_{Ni}$ ($\mathrm{M_\odot}$)& 0.016 & - & 0.01 & - & 0.056 & - & 0.02 & - \\
$E_{tot}$ ($10^{51}$ erg)& 0.8 & 0.61 & 0.8 & 1.0 & 2.2 & 1.0 & 1.45 & 1.39  \\
$E_{kin}/E_{th}$ & 1.67 & 60 & 1.67 & 19 & 2.67 & 9 & 3.14 & 14.4 \\
$\kappa$ ($\mathrm{cm^2/g}$)& 0.24 & 0.4 & 0.23 & 0.4 & 0.13 & 0.4 & 0.2 & 0.4 \\
\hline 
\hline 
\end{tabular}                                            
\end{table*}

\subsection{Comparison with other models}
\subsubsection{Normal Type IIP SNe}
For Type IIP SNe only the core parameters were available in the literature. So, in the subsection we only use the parameters for this inner region to compare with those from other models.

For SN 2004et, 2005cs, 2009N, 2012A, and 2012aw the comparison between the results of our semi-analytical light curve model and the parameters of other hydrodynamic calculations was already presented in a previous paper \citep{Nagy}.

For SN 2003hn neither hydrodynamic calculations nor analytic models are available in the literature at present. 

The radius of the progenitor star of SN 2013ej was found to be in the range of $(2.8 - 4.2)\cdot 10^{13}$ cm with a simple analytic function by \cite{Valenti}. \cite{Fraser} estimated the progenitor in the mass range of 8 - 15.5 M$_\odot$ from the luminosity on the pre-explosion images. Hydrodynamic calculations for SN 2013ej was published by \cite{Huang}, assuming a $^{56}$Ni mass of $0.02 \pm 0.01$ M$_{\odot}$. Their model was computed by semi-analytic and radiation-hydrodynamical simulation as well, which yields a total energy of $(0.7 - 2.1)\cdot$ $10^{51}$ erg, an initial radius of (1.6 - 4.2) $\cdot$ $10^{13}$ cm, and an envelope mass of 10.4 - 10.6 M$_\odot$. More recently, another semi-analytical model calculations were 
given by \citet{Bose15} which resulted in $M_{ej} = 12 \pm 3$ M$_{\odot}$, 
$R_0 = (3.1 \pm 0.8)\cdot$ $10^{13}$ cm, $E_{tot} \sim 2.3$ foe and $M_{Ni} = 0.02 \pm 0.002$ M$_{\odot}$.
These parameters are very similar to our results listed in Table \ref{table:13ej}. 

\begin{table}[!ht]
\caption{Model parameters for SN 2013ej} 
\label{table:13ej} 
\centering                  
\begin{tabular}{l c c c c c c}          
\hline
\hline \\                      
Parameter &\multicolumn{2}{c}{Literature} & This paper\\
& B15\tablefootmark{1} & H15\tablefootmark{2} & \\ 
\hline  \\                         
$R_0$ ($10^{13}$ cm)& 3.1  & 2.2 & 2.9\\
$M_{ej}$ ($\mathrm{M_\odot}$)& 12.0  & 10.5 &  10.0\\
$M_{Ni}$ ($\mathrm{M_\odot}$)& 0.018  & 0.02 & 0.02\\
$E_{tot}$ ($10^{51}$erg)& 2.3 & 1.4 & 1.45\\
\hline 
\hline                                            
\end{tabular}
\tablebib{(1)~\citet{Bose15}; (2)~\citet{Huang}.}
\end{table}

\subsubsection{SN 1987A, a peculiar Type IIP SN}
SN 1987A was a peculiar Type IIP explosion event. The pro\-ge\-nitor of SN 1987A was a blue supergiant star (BSG), named Sk -69 202 (for a review, see, e.g., \cite{Arnett89}). The BSG progenitor indicates that during the stellar evolution the mass-loss was significant. Thus, in this case we assume that our shell model refers to a mass-loss event shortly before the supernova explosion, which may create a low-mass circumstellar medium (CSM) around the progenitor. So, for fitting SN 1987A we use only the estimated parameters of the core configuration to compare with other models. 

The fundamental quantities of SN 1987A were calculated with numerous different models by several authors \citep{Arnett, Blinnikov, Nomoto, Shigeyama, Utrobin}. The LC of this event was analyzed by \cite{Blinnikov} with the hydrodynamic code STELLA. That calculation assumed that the ejected mass was $14.7 \mathrm{M_\odot}$, the kinetic energy was $1.1 \pm 0.3$ foe and the radius of the progenitor star was about $3.4 \cdot 10^{12}$ cm. The hydrodynamic calculation of \cite{Shigeyama} estimated that the radius of the progenitor star should have been about $(1 - 3) \cdot 10^{12}$ cm. In that publication an ejected mass of 7 - 10 $\mathrm{M_\odot}$ and an explosion energy of 2 - 3 foe was found. Another hydrodynamic model was presented by \cite{Nomoto}, which predicted 11.3 $M_\odot$ for ejected mass, 0.07 $\mathrm{M_\odot}$ for the initial nickel mass and 1.5 foe for the explosion energy. The LC of SN 1987A was also analyzed by \cite{Utrobin}, and their hydrodynamic model presents an ejected mass of $18 \mathrm{M_\odot}$, a kinetic energy of 1.5 foe, a nickel mass of 0.077 $\mathrm{M_\odot}$ and a radius of $2.44 \cdot 10^{12}$ $cm$. The physical properties of this supernova was also calculated by \cite{Arnett} with the values of $R_0 = 1.05 \cdot 10^{12}$ cm, $M_{ej} = 7.5$ $M_\odot$, $E_{tot} \approx 1.5$ foe, and $M_{Ni} = 0.075 \pm 0.015$ M$_\odot$. The fundamental parameters of SN 1987A were also inferred by \cite{Imshennik}. They found that the radius of the progenitor was $(1.8 - 2.8) \cdot 10^{12}$ cm, the kinetic energy was 1.05 - 1.2 foe, the ejected mass was 15.0 $\mathrm{M_\odot}$ and the nickel mass was 0.071 - 0.078 $\mathrm{M_\odot}$. The comparison of all of these parameters with the ones from our LC fitting code can be found in Table~\ref{table:87A}.

From Table~\ref{table:87A} it is seen that the numerous efforts for mo\-deling the LC of SN 1987A led to quite different physical para\-meters. This is especially true for the ejected mass, where a factor of 2 disagreement between the results from the early and modern hydrodynamic codes can be found. The ejected mass from our semi-analytic code, as well as the other parameters, are consistent with these hydrodynamic results. It is true that the semi-analytic code tend to predict somewhat lower ejected masses than some of the modern hydrocodes, but the difference is not higher than the differences between the results from various hydrodynamic calculations. 

\begin{table*}[!ht]
\caption{Model parameters for SN 1987A} 
\label{table:87A} 
\centering                  
\begin{tabular}{l c c c c c c c c}          
\hline
\hline \\                      
Parameter &\multicolumn{7}{c}{Literature} & This paper\\
& S87\tablefootmark{1} & N87\tablefootmark{2} & AF89-1\tablefootmark{3} & AF89-2\tablefootmark{3}& IP92\tablefootmark{4} & B00\tablefootmark{5} & UC05\tablefootmark{6} & \\ 
\hline  \\                         
$R_0$ ($10^{12}$ cm)& 2.0 & 4.5 & 1.05 & 0.75 & 2.3 & 3.4 & 2.44 & 2.9\\
$M_{ej}$ ($\mathrm{M_\odot}$)& 8.5 & 11.3 & 7.5 & 15 & 15 & 14.7 & 18 & 8.6\\
$M_{Ni}$ ($\mathrm{M_\odot}$)& < 0.1 & 0.07 & 0.075 & 0.75 & 0.075 & 0.078 & 0.077 & 0.069\\
$E_{kin}$ ($10^{51}$erg)& - & - & 1.0 & 2.0 & 1.1 & 1.1 & 1.5 & 1.4\\
$E_{tot}$ ($10^{51}$erg)& 2.5 & 1.5 & 1.5 & 3.0 & - & - & - & 1.52\\
$t_d$ (day)& 63.4 & 78.5 & 64.1 & 90.6 & 118.7 & 103.4 & 111.4 & 63.6\\
\hline 
\hline                                            
\end{tabular}
\tablebib{(1)~\citet{Shigeyama}; (2)~\citet{Nomoto}; (3)~\citet{Arnett}; (4)~\citet{Imshennik} (5)~\citet{Blinnikov}; (6)~\citet{Utrobin}.}
\end{table*}

Although some major differences can be noticed between the published results (especially in $M_{ej}$), the calculated parameters are in the same order-of-magnitude, and our estimates are entirely consistent with these parameter ranges. However, it should be kept in mind that in analytic models, which use the constant opacity approximation, the correlation between $\kappa$ and $M_{ej}$ (Sec. 3.2) has a significant effect on the estimated $M_{ej}$. Thus, in these models the ejected mass may be constrained only within a factor of two, if the assumed opacity is chosen between 0.1 - 0.3 cm$^2$/g.

In the last row of Table~\ref{table:87A} we show the mean light curve time-scales, which is equal to the effective diffusion time-scale, $t_d = \sqrt{2 t_a t_h}\ \approx 1.89\cdot 10^{-11} (\kappa^2 M_{ej}^3 / E_{kin})^{1/4}$ days \citep{Arnett80, Arnett89}, as inferred from the parameters of the different models (assuming $\kappa \sim 0.2$ cm$^2$g$^{-1}$ for those that were not computed with constant opacity). This quantity is roughly proportional to the risetime to the LC peak (or the length of the plateau in Type IIP SNe) in the constant-density model. Since at least half of the models (S87, N87, B00, UC05) listed in Table~\ref{table:87A} were not computed with the constant density approximation, this parameter is less applicable to those models, but generally it gives an good estimate of the {\it relative} peak times of the different models. 

It is interesting that the models predicting more ejecta mass ($M_{ej} > 10$ $M_\odot$) all have 
$t_d > 70$ days, while the models with less massive ejecta consistently have peak times less than 70 days. Although for models with recombination taken into account the actual LC peak time is somewhat longer than $t_d$ \citep[e.g.,][]{Arnett}, it is true that the models having longer $t_d$ tend to show later peaks than the models with shorter $t_d$. Comparing the $t_d$ values in Table~\ref{table:87A} with the observed peak time of SN~1987A ($t_d \sim 90$ days, Fig.~\ref{fig:IIP_1} upper panel), it is seen that for the models with lower masses ($M_{ej} <10$ $M_\odot$) the inferred $t_d$ parameters are consistent with the observed peak time.

Concerning the more massive models, the AF89-2 model ($M_{ej} = 15$ $M_\odot$, $t_d = 91$ d) is still more-or-less consistent with the observed LC peak, but the other three models (IP92, B00, UC05) tend to predict longer peak times than observed. This may not be surprising as far as the B00 and UC05 models are concerned, because those are radiative hydrodynamical models, for which the simple analytic estimate from the constant density model may not be applicable. However, the model of IP92 \citep{Imshennik} is also a semi-analytic model which uses similar approximations to our one in this paper. Thus, it is not clear why this model is able to fit the observed peak of the LC of SN~1987A, while it has the longest predicted $t_d$ among all models listed in Table~\ref{table:87A}. When compared with  the AF89-2 model \citep{Arnett}, which also has $M_{ej}=15$ $M_\odot$, it is seen that the latter one has a factor of two higher kinetic energy than the IP92 model. Since higher $E_{kin}$ causes the computed LC to peak earlier, the AF89-2 model with higher $E_{kin}$ looks more plausible for fitting SN~1987A than the IP92 model. 

It is concluded that the cause of the wide range of the predicted physical parameters listed in Table 10, in addition to the already discussed issues such as the constant opacity approximation and the opacity - mass correlation, is supposedly the strong correlations between practically all parameters involved in the LC modeling \citep[e.g.,][]{Arnett, Imshennik, Nagy}. If one, for example, assumes $M_{ej} \sim 15$ $M_\odot$ ejecta mass, it is possible to find e.g. a kinetic energy that is physically realistic and the model gives a good fit to the observed LC. The same is true for $M_{ej} < 10$ $M_\odot$,
which needs lower $E_{kin}$, but still in the ballpark, to get almost the same fitting.
This is clearly a caveat of SN LC modeling, which should be kept in mind when interpreting the SN parameters inferred from pure LC fits.

\section{Expansion velocity of CCSNe}
An additional test of the LC models can be made via the comparison of the expansion velocities derived 
from the best-fitting LC models with the observed ones. We apply Eq. 11 to calculate the expansion velocities of the SNe under study, and these values can be compared with the available velocities from spectroscopic measurements collected from literature. Since the expansion velocity (the maximum velocity in the applied semi-analytic model) cannot be measured directly, we use the observed photospheric velocities ($v_{ph}$) from early and late-time spectra as a proxy for the expansion velocities \citep[see also ][]{Wheeler}. 
 
As a first approximation, we compare the expansion velocities of the envelope ($v_{shell}$) with
the earliest $v_{ph}$ values, when the photosphere is likely in the outer envelope 
\citep[see e.g.][]{Moriya}. For the core expansion velocities ($v_{core}$), we use the late-phase
$v_{ph}$ values measured around the middle of the plateau phase. Table~\ref{table:vel1} and ~\ref{table:vel2} list these data for the envelope and the core, respectively. 

\begin{table}[!h]
\caption{Velocities of the outer envelope of the modeled SNe}
\label{table:vel1}
\centering                  
\begin{tabular}{l c c l}          
\hline
\hline \\                      
Supernova & Model $v_{shell}$ & Observed $v_{ph}$ & Reference\\
  & [$10^3$ km s$^{-1}$] & [$10^3$ km s$^{-1}$] & \\  
\hline  \\                         
SN 1987A & 28.3 & $\sim$ 30.0 & 1\\
SN 2004et & 15.7 & $\sim$ 14.0 & 2\\
SN 2012A & 16.2 &  $\sim$ 10.0 & 3 \\ 
SN 2012aw & 14.2 & $\sim$ 12.0 & 4 \\
\hline 
\hline                                            
\end{tabular}
\tablebib{(1)~\citet{Hans}, (2)~\citet{Utrobin09}, (3)~\citet{Tomasella}, (4)~\citet{Bose}.}
\end{table}

\begin{table}[!h]
\caption{Velocities of the inner ejecta of the modeled SNe}
\label{table:vel2}
\centering                  
\begin{tabular}{l c c l}          
\hline
\hline \\                      
Supernova & Model $v_{core}$ & Observed $v_{ph}$ & Reference\\
  & [$10^3$ km s$^{-1}$] & [$10^3$ km s$^{-1}$] & \\  
\hline  \\                         
SN 1993J & 13.7 & $\sim$ 14.0 & 1\\
SN 2011fu & 13.5 & 13.0 - 14.0 & 2 \\
SN 1987A & 5.22 & 2.0 - 3.5 & 3 \\
SN 2004et & 4.25 & 3.3 - 3.6 & 4 \\
SN 2005cs & 2.58 &  1.0 - 1.5 & 5 \\
SN 2009N & 3.33 & 2.5 - 3.0 & 6\\
SN 2012A & 3.23 & $\sim$ 3.0 & 7 \\ 
SN 2012aw & 3.65 & 3.4 - 3.6 & 8 \\
SN 2013ej & 4.28 & $\sim$ 4.6 & 9\\
\hline 
\hline                                            
\end{tabular}
\tablebib{(1)~\citet{Bartel}, (2)~\citet{Morales}, (3)~\citet{Larsson}, (4)~\citet{Maguire}, (5)~\citet{Pastorello}, (6)~\citet{Takats},(7)~\citet{Tomasella}, (8)~\citet{Bose}, (9)~\citet{Huang}.}
\end{table}

Note that the explosion velocities derived from LC modeling are always somewhat uncertain 
due to the correlation bet\-ween the ejected mass and the kinetic energy \citep[see e.g.][]{Nagy}.
This may result in significant systematic offset in the model velocities. Thus, 
we are able to estimate the expansion velocity only within a factor of $\sim$ 2. 
Considering these circumstances, the range of the observed velocities in Table~\ref{table:vel1} and ~\ref{table:vel2} are in reasonable agreement with our model values. 

\begin{figure}[!ht]
\includegraphics[width=9cm]{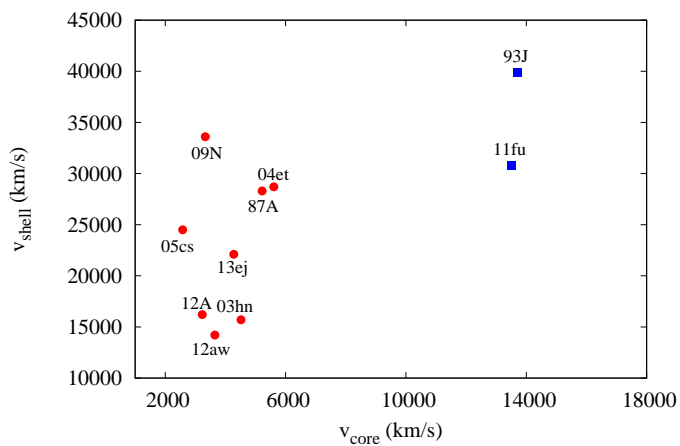}
\caption{Expansion velocities of Type IIb (blue squares) and Type IIP (red dots) supernovae. }
\label{fig:vel}
\end{figure}

We also check the correlation between $v_{core}$ and $v_{shell}$ for both Type IIb and Type IIP SNe. As Fig.~\ref{fig:vel} shows, these two types of explosion events appear to be separated in velocity space. Type IIb SNe have larger velocities in both regions, while for Type IIP SNe only $v_{shell}$ may reach higher values. The differences in $v_{core}$ can be explained by the difference between the fitting parameters, especially $M_{ej}$ and $E_{kin}$. Because the kinetic energies  are in the same order of magnitude, while the ejected masses for Type IIP-s are 10 times lager than for Type IIb-s, the inferred velocities for the latter are lower, in accord with simple physical expectations.       

\section{Discussion and conclusions}

We showed above that for Type IIP SNe a two-component ejecta configuration can be appropriate to model a double-peaked light curve with a semi-analytic diffusion-recombination code. This ejecta configuration 
is similar to the one used for Type IIb SNe assuming an inner, denser core and a hydrogen-dominated outer envelope, although for Type IIb-s the envelope is much more extended than the core, unlike in
Type IIP models.

There are a number of caveats in the simple diffusion-recombination model applied in this paper, 
such as the assumption of constant opacity in the ejecta, which naturally limits the accuracy of the
derived physical parameters. Although the Thompson-scattering opacity approximation simplifies the
equations, the used opacity lose its physical meaning and becomes only a fitting parameter. But the average opacities from the SNEC hydrodynamic model show adequate agreement with the frequently used opacities in literature and also with our applied $\kappa$-s. Another limitation is the uncertainty of the explosion date, which may cause 5 - 50 $\%$ relative errors in the derived phy\-sical parameters. Despite of these issues, by using simple semi-analytic models one can estimate the initial parameters of the progenitor and the SN with an order-of-magnitude accuracy. 

The final results in Section 5 and Section 6 show that Type IIP SNe have the most extended envelopes having initial radii of $\sim 10^{13}$ cm, which is in good agreement with the radii of their presumed RSG progenitors. For these events the ejected mass of the inner region ($M_{core}$) is about an order-of-magnitude larger than that of Type IIb SNe, which is consistent with the diffe\-rent physical state of their progenitors \citep[e.g.][]{Woosley, Heger}. 


In the diffusion approximation it is important that the contributions from the two different components should be well se\-pa\-ra\-ted. To fulfil this requirement, the diffusion time-scale of the photons in the outer envelope must be much lower than that of the inner core.
If this is so, the first LC peak can be explained as mostly due to the adiabatic cooling of the shock-heated H-rich envelope, where there is no energy input other than the thermal energy deposited by the initial shock wave. Thus, at very early phases the contribution of the outer envelope  dominates the luminosity. 
On the contrary, the second LC peak is powered by radioactive decay of $^{56}$Co and the recombination of H. The e\-ner\-gy input from these mechanisms only diffuses through the inner core before escaping as observable radiation, because at later epochs the outer envelope already became strongly diluted, thus, its effect on the light curve is negligible (Fig.~\ref{fig:IIb}). 

From the model parameters presented in Table~\ref{table:IIb} and Table~\ref{table:IIP} the effective diffusion time-scale $t_d$  can be inferred, which weakly depends on the density profile. For the inner core this parameter also represents the rise time and the LC width of Type IIb and Type IIP SNe, respectively. The typical calculated $t_d$ values for Type IIP SNe are $\sim$ 80-100 days, while for Type IIb they are $\sim$ 20 days, which are in reasonable agreement with visually specified LC properties.  
Also, it can be shown that $t_d$ in the core is at least 2 times higher than in the envelope for both Type IIP and Type IIb compositions. 
Thus, the two-component models given in Table~\ref{table:IIb} and \ref{table:IIP} are consistent with the assumption that the photon diffusion equations in the two components can be solved 
separately.  

Note, however, that the Type IIP models may not be entirely self-consistent, because in those
cases the radius of the envelope is only 2 times higher than that of the core. In this 
case neglecting the effect of the radioactive heating in the outer envelope may not be
fully justifiable. This means that the inferred envelope parameters in Table~\ref{table:IIP}
are much more uncertain than the parameters of the core. In spite of this uncertainty, the 
envelope parameters occupy nearly the same regime as the CSM masses in 
the hybrid ejecta+CSM models calculated by \citet{Moriya}. 
As far as the Type IIb models are concerned, the parameters in Table~\ref{table:IIb} are also
in good agreement with the results by \citet{Nakar1} who estimated the envelope masses and the
maximum radii of the core in stripped-envelope SNe.

The two-component semi-analytic model presented in this paper may be a useful tool for 
deriving order-of-magnitude estimates for the basic parameters of Type IIP and IIb
SNe, which can be used to narrow the parameter regime in more detailed si\-mu\-lations.
The model can predict reasonable parameters for both the inner core and the outer
envelope. It can fit the entire LC starting from shortly after shock breakout
throughout the end of the plateau and extending into the nebular phase. 

An obvious advantage of the semi-analytic code with respect to the more computationally intense hydrodynamic codes is the execution time, which is $\sim 2$ minutes on a Core-i7 CPU, compared to $\sim 10$ hours needed to complete a model LC with SNEC (not to mention the more sophisticated hydrocodes that may require orders of magnitude longer time on supercomputers). 

However, it should be kept in mind that some of the parameters of the semi-analytic code, like $\kappa$, $M_{ej}$ and $E_{kin}$, are correlated, thus, only their combination are constrained by the data. Moreover, the constant $\kappa$ opacity may not have the correct physical meaning, thus, it should be considered only as a fitting parameter rather than a true representation of the "true" opacity in the SN atmosphere. With this limitation, the results from the two-component LC model are consistent with current state-of-the-art
calculations for Type II SNe. So, the estimated fitting parameters can be used for preliminary studies prior to more complicated hydrodynamic calculations.

The source code  of the program applied for calculating the semi-analytic models in this paper is available at {\tt http://titan.physx.u-szeged.hu/$\sim$nagyandi/LC2}.


\begin{acknowledgements}
We are indebted to an anonymous referee, who provided very useful comments and suggestions from which we have indeed learned a lot, and which led to significant improvement of this paper. This work has been supported by Hungarian OTKA Grant NN 107637 (PI Vinko). 
\end{acknowledgements}

\end{document}